# A New Approach to Special Relativity and The Universe

Yang-Ho Choi

**Department of Electronic and Telecommunication Engineering**
**Kangwon National University**
**Chunchon, Kangwon-Do, 200-701, South Korea**

**Abstract:** A new approach to special relativity is presented which introduces coordinate systems with imaginary time axes, observation systems, and coordinate bases. The employed space-time space lies in a complex Euclidean space (CES), which is an extension of the conventional real Euclidean space to complex numbers. In CES, the space and time axes of each coordinate system are perpendicular to each other and relative motion with a constant velocity results in the rotation of one coordinate system with respect to the other. The observation system is a set of observation points of all observers in an inertial system that uses coordinate systems with the same basis. Simultaneous events in one observation system appear simultaneously in another one as well. In other words, the absolute simultaneity is shown in the observation systems. Moreover, the observation systems enable us to clearly explain the phenomena of time dilation and length contraction. The reference observation system is the one in which time runs faster than in the other system. In relation to the transformation, the coordinates of events in the latter are dependent on the former. The conventional theory, which is based on the principle of relativity and stays at the level of coordinate systems, has made the true space-time space disappear. The true space-time space is the reference inertial system with respect to every inertial system, and time in the true space runs fastest. Assuming that the true space and time exist, the application of the new approach to the universe provides surprising results. It shows that the universe is expanding, which is equivalent to time creation. Moreover, the light speed in the true space is the expansion speed of the universe. The Hubble constant is derived based on the expanding universe model. The twin paradox problem has been resolved with the true time and space found.

(**Keywords:** Special relativity, Complex Euclidean space, Observation system, Simultaneity, True space and time, Expansion of the universe)

# 1. Introduction

Einstein's formulation of special relativity [1–5] shows a transformation between two coordinate systems, which is equal to the Lorentz transformation, when there is relative motion between them with a constant velocity. However, certain contradictions and paradoxes associated with the conventional theory of special relativity have not been clearly resolved since Einstein's historic enunciation in 1905. According to the conventional theory, simultaneity does not hold from one coordinate system to another, and it introduced the concept of the relativity of simultaneity. The relativity of simultaneity implies that the order of event occurrence in one inertial system is different from that in another system. Hence, past and future events in one system become present events in another, which seems very unnatural. The conventional theory has made the true space-time space disappear. Everything is only relative, coordinate systems being equivalent in relativity. In fact, the conventional theory has been staying at the level of coordinate systems since 1905. At the coordinate system level, any differences between two inertial systems cannot be seen except that one is moving in a direction and the other in the opposite direction. The problems of the simultaneity and the true space cannot be properly dealt with at the coordinate system level. Even the coordinate system employed is difficult to use for the investigation on special relativity.

In this paper, a new approach is presented which significantly differently deals with special relativity, introducing coordinate systems with imaginary time axes, observation systems, and coordinate bases. As the coordinate systems use imaginary time axes, the space-time space employed by the new approach lies in a complex Euclidean space (CES), which is an extension of the conventional real Euclidean space to complex numbers. In the Minkowski space [2–5], which lies in a real number space, the space and time axes of the coordinate system for one observer in constant motion relative to another are not perpendicular to each other, while in CES, the space and time axes of each coordinate system are perpendicular. The constant relative motion in CES results in the rotation of one coordinate system with respect to the other and the rotation is described by a complex angle between the two coordinate systems.

The observation system is a set of observation points for all observers in a relative motion with a constant velocity. Each observer needs its own coordinate system such that events can be represented in the system with the space and time axes. The observation system also needs a coordinate system to represent the observation points. In this paper, unless indicated otherwise, 'a coordinate system', which is also called 'an inertial frame' in the literature on special relativity, refers to 'the coordinate system of an observer', not of the observation system. The term 'inertial system' is used as a generic term for a whole system in relative motion. A single inertial system includes one observation system and the coordinate systems of all observers belonging to it. The coordinate bases for the observation system and the coordinate systems which belong to the same inertial system are the same.

The time axis of a coordinate system is the set of the observation points, the observation line, which its observer observes. In the conventional theory, each line parallel to the time axis as well as the time axis is regarded as the observation line of an observer belonging to the same inertial system. Unless the coordinate system is equal to its observation system, however, the other lines except the time axis are not observation lines and the coordinate system has only one observation line, the time axis, for one observer. Therefore we cannot properly deal with the observation-related problems involving two or more observers in one inertial system by using the coordinate system only. It is the reason why the coordinate-system-based conventional approach could not have provided clear explanations on simultaneous events. The introduction of the observation systems allows us to find the observation lines of all observers. Events occur simultaneously in every observation system, which leads to the absolute simultaneity. The phenomena of time dilation and length contraction are clearly explained in the observation systems without any contradiction.

The principle of relativity in the conventional approach has made the true space and time disappear. However, the new approach paves the way to find the true space and time. As far as the coordinate transformation is concerned, only one observation system can be identical to the coordinate system of an observer. The equality in relativity no longer holds in the observation systems. The reference observation system is the one in which time runs faster than in the other system. The latter is dependent on the former in terms of the coordinate transformation. These lead us to know what the true space-time space is. The true space is the reference inertial system where time runs faster than in any other inertial system and its observation system is the same as the coordinate system of an observer. Under the assumption that the true space and time exist, the application of the new approach to the universe surprisingly shows that the universe expands, thereby creating time. Additionally, the light speed in the true space is the expansion speed of the universe. The Hubble constant [6] is derived using the expanding universe model. It seems that the twin paradox has remained unsolved so far, though some explanations were given [2, 5]. The twin paradox problem has been resolved with the true space-time space found.

Einstein adopted two axioms in the derivation of the formulation of special relativity. One is that physical laws can be represented in the same forms in any inertial system. The other is that the light speed is constant in every inertial frame. The principle of relativity includes the same physical laws. A sort of the Lorentz transformation can be attained using only the relativity principle without the axiom of the constant light speed [1]. However, in the approach, an additional condition is required to determine an unknown parameter. An invariant speed can be given for the condition. If the invariant speed is the light speed, it is equivalent to employing the axiom of the constant light speed. In this paper, the light speed is assumed to be constant in every coordinate system, as in the conventional approach.

Following this Introduction, Section 2 presents the new coordinate systems, from which Einstein's formulation of special relativity is derived. The problem of simultaneity is treated, introducing the observation systems. Moreover, the true space-time space is described together with coordinate bases. In Section 3, the new approach is applied to the universe to find the true space-time space. The Hubble constant is derived. In Section 4, the twin paradox problem is dealt with, using results obtained in Sections 2 and 3, and is followed by Conclusions.

## 2. Special Relativity, Coordinate Systems and Observation Systems

*2.1 Special Relativity and Complex Coordinate Systems*

Consider a coordinate system $S = (\tau, x)$ shown in Fig. 1, which has the imaginary time axis of $\tau = ict$ where $i = \sqrt{-1}$, $c$ is the light speed, and $t$ denotes time. The Minkowski space uses the real time axis, which is set at $ct$, considering that electromagnetic waves propagate with the light speed. Then a distance $d_1$ form the origin to a point $p_{S,1} = (ct_1, x_1)$ is given by $d_1 = \sqrt{x_1^2 - ct_1^2}$. In the new coordinate system with the imaginary time axis, the distance $d_1$ is expressed as $d_1 = \sqrt{\tau_1^2 + x_1^2}$, and then the space which the new coordinate system employs can be considered to be a Euclidean space extended to complex numbers. We call it a complex Euclidean space. For a point $p_{S,a} = (\tau_a, x_a)$, we will also use the vector notation $\boldsymbol{p}_{S,a} = [\tau_a, x_a]^T$ where $T$ stands for transpose. In CES, the distance $d(p_{S,a})$ and the length $l(p_{S,a})$ of $\boldsymbol{p}_{S,a}$ are defined as

$$d(\tau_a, x_a) = d(\boldsymbol{p}_{S,a}) = \|\boldsymbol{p}_{S,a}\| = \sqrt{\tau_a^2 + x_a^2} \tag{1}$$

$$l(\tau_a, x_a) = l(\boldsymbol{p}_{S,a}) = |d(\tau_a, x_a)| \tag{2}$$

where $|\zeta|$ denotes the magnitude of a complex number $\zeta$. The distance $d(p_{S,2}, p_{S,1})$ from $p_{S,1} = (\tau_1, x_1)$ to $p_{S,2} = (\tau_2, x_2)$ is given by

$$\begin{aligned} d(\tau_2, x_2; \tau_1, x_1) &= d(\boldsymbol{p}_{S,2}, \boldsymbol{p}_{S,1}) \\ &= \|\boldsymbol{p}_{S,2} - \boldsymbol{p}_{S,1}\| = [(\tau_2 - \tau_1) + (x_2 - x_1)]^{1/2} \end{aligned} \tag{3}$$

Imaginary quantities for time elements, under the Minkowski structure, have been used, for example, to represent distances [4, 5].

An observer $O'$ is moving along the $x$-axis with a constant velocity of $v$ with respect to $O$ who is located at $x = 0$. Suppose that the two observers meet at $x = 0$ when $\tau = 0$. At $t = t_1$, $O'$ is at the point $p_{S,1} = (ict_1, vt_1)$. As time passes, $O$ goes along the $\tau$-axis, while $O'$ follows the $\tau'$- line which is a line crossing $O'$ and $p_{S,1}$, as shown in Fig. 1, where the same symbol is used to denote both a coordinate origin and an observer. Hence, we set the $\tau'$-line as the time axis for $O'$, putting $\tau' = ict'$. Note that in Fig. 1, the $x$-axis is located perpendicular to the $\tau$-axis in the counter clockwise direction. Accordingly, the space axis, the $x'$-axis, for $O'$ is set. Though not shown in Fig. 1, the other axes for $O'$, the $y'$- and $z'$-axes, are the same as the $y$- and $z$- axes for $O$. We will not consider the $y'$- and $z'$-coordinates, if not necessary.

The angles $\theta_a$ and $\theta_1$ in Fig. 1 are complex numbers because $\tau$ and $\tau'$ are complex. For a complex $\theta$, $\cos\theta$ and $\sin\theta$ are defined as [7]

$$\cos\theta = \frac{e^{i\theta} + e^{-i\theta}}{2} \tag{4}$$

$$\sin\theta = \frac{e^{i\theta} - e^{-i\theta}}{2i}. \tag{5}$$

It is straightforward to show that $\cos^2\theta + \sin^2\theta = 1$. The point $p_{S,a}$ can be represented in polar form as

$$\boldsymbol{p}_{S,a} = \begin{bmatrix} \tau_a \\ x_a \end{bmatrix} = \begin{bmatrix} \|\boldsymbol{p}_{S,a}\|\cos\theta_a \\ \|\boldsymbol{p}_{S,a}\|\sin\theta_a \end{bmatrix}. \tag{6}$$

The point $p_{S',a}$ in $S' = (\tau', x')$ corresponding to $p_{S,a}$ is expressed as

$$\boldsymbol{p}_{S',a} = \begin{bmatrix} \tau_a' \\ x_a' \end{bmatrix} = \begin{bmatrix} \|\boldsymbol{p}_{S,a}\|\cos\theta_{a1} \\ \|\boldsymbol{p}_{S,a}\|\sin\theta_{a1} \end{bmatrix} \tag{7}$$

with $\theta_{a1} = \theta_a - \theta_1$. In complex numbers also, the trigonometric identities hold [7]:

$$\cos(\theta_a \pm \theta_1) = \cos\theta_a \cos\theta_1 \mp \sin\theta_a \sin\theta_1 \tag{8}$$

$$\sin(\theta_a \pm \theta_1) = \sin\theta_a \cos\theta_1 \pm \cos\theta_a \sin\theta_1. \tag{9}$$

It is easy to prove (8) and (9) using the definitions (4) and (5). Substituting (8) and (9) into (7) with the use of (6), it follows that

$$\begin{bmatrix} \tau_a' \\ x_a' \end{bmatrix} = \begin{bmatrix} \cos\theta_1\, \tau_a + \sin\theta_1\, x_a \\ -\sin\theta_1\, \tau_a + \cos\theta_1\, x_a \end{bmatrix}. \tag{10}$$

From Fig. 1,

$$\cos\theta_1 = \frac{d(\tau_1, 0)}{d(p_{S,1})} = \frac{1}{\sqrt{1 - (v/c)^2}} \tag{11}$$

$$\sin\theta_1 = \frac{d(0, x_1)}{d(p_{S,1})} = \frac{v/c}{i\sqrt{1 - (v/c)^2}} \tag{12}$$

where $p_{S,1} = (\tau_1, x_1) = (ict_1, vt_1)$. Note that for $v < c$, $\cos\theta_1$ is real while $\sin\theta_1$ is imaginary. It is easy to see by the substitution of (11) and (12) into (10) with $\tau_a = ict_a$ and $\tau_a' = ict_a'$ that

$$t_a' = \frac{t_a - x_a v/c^2}{\sqrt{1-(v/c)^2}} \tag{13}$$

$$x_a' = \frac{x_a - vt_a}{\sqrt{1-(v/c)^2}}, \tag{14}$$

which are the same as Einstein's formulations.

According to (10) with the replacement of $\theta_1$ by $\theta$, $\boldsymbol{p}_{S'} = [\tau', x']^T$ is related to $\boldsymbol{p}_S = [\tau, x]^T$ by

$$\boldsymbol{p}_{S'} = \boldsymbol{T}(\theta)\boldsymbol{p}_S \tag{15}$$

where

$$\boldsymbol{T}(\theta) = \begin{bmatrix} \cos\theta & \sin\theta \\ -\sin\theta & \cos\theta \end{bmatrix}. \tag{16}$$

The transformation matrix $\boldsymbol{T}(\theta)$ from $S$ to $S'$ sets $S'$ to be rotated by a rotation angle $\theta$ with respect to $S$ where $\theta$ is measured in the direction from the time axis to the space axis of $S$, as shown in Fig. 1. Even if $\tau$, $\tau'$, $x$, and $x'$ are changed into $ix$, $ix'$, $-i\tau$, and $-i\tau'$, respectively, (15) is the same, which implies that the space and the time have the relationship of duality. It is straightforward to see that

$$\boldsymbol{T}^T(\theta)\boldsymbol{T}(\theta) = \boldsymbol{I} \tag{17}$$

where $\boldsymbol{I}$ is an identity matrix. The distance is known to be an invariant quantity with respect to inertial coordinate systems [2–5]. We can confirm it by calculating the distance between two points $\boldsymbol{p}_{S',1} = [\tau_1, x_1]^T$ and $\boldsymbol{p}_{S',2} = [\tau_2, x_2]^T$, which is given by

$$\begin{aligned} d(\boldsymbol{p}_{S',2}, \boldsymbol{p}_{S',1}) &= \|\boldsymbol{p}_{S',2} - \boldsymbol{p}_{S',1}\| \\ &= \sqrt{(\boldsymbol{p}_{S,2} - \boldsymbol{p}_{S,1})^T \boldsymbol{T}^T(\theta)\boldsymbol{T}(\theta)(\boldsymbol{p}_{S,2} - \boldsymbol{p}_{S,1})} \\ &= \|\boldsymbol{p}_{S,2} - \boldsymbol{p}_{S,1}\| = d(\boldsymbol{p}_{S,2}, \boldsymbol{p}_{S,1}). \end{aligned}$$

In the Minkowski space, the time and space axes of $S'$, which are denoted as the $ct'$- and $x'$-axes respectively, correspond lines $x = (v/c)ct$ and $ct = (v/c)x$, respectively, in $S$. In Fig. 2, the new coordinate system which lies in CES is compared with the Minkowski coordinate system which lies in the real number space. The space and time axes of $S'$ in CES are perpendicular to each other, whereas those in the real number space are not.

Einstein derived the Lorentz transformation based on the two axioms: the constant light speed and the same physical laws. The relativity principle includes the same physical laws. A sort of the Lorentz transformation can be obtained such that the properties of the relativity, the isotropy and the homogeneity, without the axiom of the constant light speed, are satisfied [1]. The relativity means that when an observer $O''$ is moving with a velocity of $-v$ along the $x'$-axis, the coordinate system $S''$ for $O''$ should be the same as $S$. Let the rotation angle of $S''$ with respect to $S$ be $\theta_2$. The transformation matrix $T(\theta_2)$ is given by

$$T_{S''|S}(\theta_2) = T_{S''|S'}(-\theta_1) T_{S'|S}(\theta_1) = I \tag{18}$$

where subscripts in transformation matrices are used to indicate the transformation direction, for example, $T_{B|A}(\theta)$ representing the transformation from $A$ to $B$. Equation (18) leads to $\theta_2 = 0$, and $S'' = S$. It is easy to see that the other properties are also satisfied. The velocity addition law can be easily obtained by the rotation relationship. From (11), $\cos\theta_k = 1/\sqrt{1-(v_k/c)^2}$ and $\sin\theta_k = -i(v_k/c)/\sqrt{1-(v_k/c)^2}$ where $k = 1, 2, 3$, and $\theta_3 = \theta_1 + \theta_2$. Using the trigonometric identity, $\cos(\theta_1 + \theta_2)$ is expressed as

$$\cos(\theta_1 + \theta_2) = \frac{1+(v_1/c)(v_2/c)}{\sqrt{1-(v_1/c)^2}\sqrt{1-(v_2/c)^2}}. \tag{19}$$

Equation (19) is equal to $\cos\theta_3 = 1/\sqrt{1-(v_3/c)^2}$, and from the equality, we have

$$v_3 = \frac{v_1 + v_2}{1+(v_1 v_2/c^2)}, \tag{20}$$

which corresponds to the known velocity addition law [1−5]. It is seen from (20) that the light speed $c$ is an invariant speed.

In general, the correct position in $S'$ of $p_{S',a}$, which can be found from (13) and (14), is different from that in $S$ of $p_{Sa}$, though in Fig. 1, for convenience, it is drawn as if they are at the same position. In $S$, a line crossing points $O$ and $p_{S,a} = (\tau_a, x_a)$ is expressed as

$$ix = \eta\tau \tag{21}$$

where the real number $\eta$, the slope of the line, is $\eta = ix_a/\tau_a$. Using (11), (12) and (15), we obtain the representation in $S'$ of line (21) as

$$ix' = \eta'\tau' \tag{22}$$

where

$$\eta' = \frac{\eta - v/c}{1 - \eta v/c} \tag{23}.$$

The quantities $\eta$ and $\eta'$ are equivalent to velocities normalized with respect to $c$. When $\eta = v/c$, $\eta'$ becomes zero, which implies that moving with a velocity of $v$ in $S$ is equivalent to being at rest in $S'$. When $\eta = 1$, $\eta'$ also is equal to one, which indicates the constancy of the light speed in every inertial frame. Fig. 3 shows the lines in $S'$ corresponding to the lines in $S$ for $\eta = 0$, $v/c$, $1$, and $\infty$. Note that each time axis overlaps the corresponding line. In other words, the $\tau$- and $\tau'$-axes correspond to $ix' = -i(v/c)\tau'$, which is the representation in $S'$ of the $\tau$-axis, and $ix = i(v/c)\tau$, which is the representation in $S$ of the $\tau'$-axis, respectively. The other lines except these two axes are rotated when being transformed from one coordinate system to another.

*2.2. Observation Systems and Simultaneity*

It is the observation point that we are concerned in the transformation (15) about. The point $p_{S',a}$ in Fig. 1 can be observed by, for example, an observer $O''$ whose $\tau''$-axis is the line crossing two points $O$ and $p_{S,a}$. Only the $\tau'$-axis is the set of observation points, the observation line, of $O'$. When $\tau = \tau' = 0$, the observer $O'$ meets $O$, and an event at each point on the $x$-axis occurs. When $\tau' = 0$, the events at the points on the $x$-axis are also observed by the observers in the inertial system $S'$ who are at the respective points. The term 'inertial system' is used as generic terminology, as mentioned in Section 1. For convenience, symbol $S'$ was also used to denote the inertial system though it is different from the coordinate system $S'$ of the observer $O'$. To avoid confusions when the same notation, say $S'$, denotes the inertial system as well, it will be expressed by inserting the phrase 'inertial system' as 'inertial system $S'$'. The coordinate system $S' = (\tau', x')$ has been established in accordance with the observation of the observer $O'$ who is at $x' = 0$. It is just for the observer $O'$, and cannot be applied to the observations of other observers. To obtain the observation points for an observer $O'^{(d')}$ who is at $x' = d'$, it is necessary to introduce the coordinate system $S'^{(d')} = (\tau'^{(d')}, x'^{(d')})$. The $\tau'^{(d')}$- and $x'^{(d')}$-axes of $S'^{(d')}$ are related to those of $S'$ by

$$\tau'^{(d')} = \tau' \tag{24}$$

$$x'^{(d')} = x' - d. \tag{25}$$

For $d' = 0$, notations are used without superscript '(0)', like $\tau'$ and $x'$ rather than $\tau'^{(0)}$ and $x'^{(0)}$. Fig. 4 shows the coordinate systems $S'^{(d')}$ together with $S'$. Line $x' = 0$, the $\tau'$-axis of $S'$, is the set of the events points that $O'$ observes. Line $x'^{(d')} = 0$, the $\tau'^{(d')}$-axis of $S'^{(d')}$, is the set of the events points that $O'^{(d')}$ observes. Note that a point $(\tau'^{(d')}, 0)$ corresponds to $(\tau', d')$ in $S'$. We define the set of the observation points for all observers in the inertial system $S'$ as

$$\mathsf{S'} = \{(\tau', d') \mid d', t' \in \Re\} \tag{26}$$

where a point $(\tau', d')$ is, according to (24) and (25), the representation in $\mathsf{S'}$ of $(\tau'^{(d')}, 0)$, a set $\{(\tau', d') \mid t' \in \Re\}$ is the observation line of $O'^{(d')}$, and $\Re$ is the set of real numbers. The set $\mathsf{S'}$ consists of all the observation points that each observer $O'^{(d')}$ observes. We call $\mathsf{S}'$ the observation system of the inertial system $S'$. The events at the same time in $\mathsf{S}'$ occur simultaneously. The observation system $\mathsf{S}$ of the inertial system $S$ is similarly defined. Since, as far as the coordinate transformation is concerned, coordinates in one coordinate system are determined relative to the other coordinate system according to (15), one coordinate system, say $S$, must be identical to its observation system $\mathsf{S}$ so that the event $e_S(\tau_a, x_a)$ at a point $(\tau_a, x_a)$ in $S$ is equal to the event $e_{\mathsf{S}}(\tau_a, x_a)$ at the point $(\tau_a, x_a)$ in $\mathsf{S}$.

Since $\tau' = 0$ when $\tau = 0$, $\tau'^{(d')}$ $(=\tau')$ is also zero at $\tau = 0$. The observer $O'^{(d')}$ meets an observer $O^{(d)}$ while $O'$ meets $O$ when $\tau' = \tau = 0$. The relationship between $S'^{(d')}$ and $S^{(d)}$ is the same as that between $S'$ and $S$, and is shown in Fig. 1 with necessary notation changes such as the replacements of $O$ and $O'$ by $O^{(d)}$ and $O'^{(d')}$, respectively. According to (15) with the notations changed, a point $\boldsymbol{p}_{S^{(d)}} = [\tau^{(d)}, x^{(d)}]^T$ is represented in $S'^{(d')}$ as

$$\boldsymbol{p}_{S'^{(d')}} = [\tau'^{(d')}, x'^{(d')}]^T = \boldsymbol{T}(\theta)\boldsymbol{p}_{S^{(d)}} \tag{27}$$

where $\tau^{(d)} = \tau$ and $x^{(d)} = x - d$.

Let $e_S(\tau_a, x_a) = e_{\mathsf{S}}(\tau_a, x_a)$, which means $S = \mathsf{S}$. Line $\tau = \tau_a$ in $\mathsf{S}$ is a set of simultaneous points. Simultaneous points mean points where events occur simultaneously. We wish to find how line $\tau = \tau_a$ is represented in the observation system $\mathsf{S}'$. The observation point in $S'^{(d')}$ when $\tau = \tau_a$ can be found by eliminating $x^{(d)}$ in (27) with $\tau = \tau_a$ and $\theta = \theta_1$, which gives

$$\cos\theta_1 \tau'^{(d')} - \sin\theta_1 x'^{(d')} = \tau_a. \tag{28}$$

The observation point $p_{S',(\tau_a',0)} = (\tau_a', 0)$ of $O'$ can be obtained by substituting $x'^{(d')} = 0$ into (28) with $d' = 0$, which gives

$$\cos\theta_1 \tau_a' = \tau_a. \tag{29}$$

Substituting $x'^{(d')} = 0$ into (28), we have the observation point $p_{S'^{(d')},(\tau_a'^{(d')},0)} = (\tau_a'^{(d')}, 0)$ of $O'^{(d')}$ as

$$\cos\theta_1 \tau_a'^{(d')} = \tau_a. \tag{30}$$

From (29) and (30), $\tau_a'^{(d')} = \tau_a'$, which indicates that the $\tau'$-coordinates of the points in $\mathsf{S}'$ corresponding to line $\tau = \tau_a$ in $\mathsf{S}$ are all equal to $\tau_a'$. Hence, the set of the points in $\mathsf{S}'$ corresponding to $\tau = \tau_a$ is expressed as

$$\tau' = \tau_a'. \tag{31}$$

Equation (31) implies that simultaneous events in one inertial system are also simultaneous ones in the other inertial system, and thus events occur simultaneously in every inertial system, which leads to the absolute simultaneity. Recall that $S = \mathsf{S}$. Line $\tau' = \tau_a'$ in $\mathsf{S}'$ and line $\tau = \tau_a$ in $\mathsf{S}$ are illustrated in Fig. 5.

The point in $\mathsf{S}$ corresponding to the point $p_{\mathsf{S}',(\tau_a',0)} = (\tau_a',0)$ in $\mathsf{S}'$ will be denoted as $p_{\mathsf{S},(\tau_a,x_0)} = (\tau_a, x_0)$. The point $p_{\mathsf{S},(\tau_a,x_0)}$ is on line $x = \tan\theta_1 \tau$, which overlaps the $\tau'$-axis, where $\tan\theta_1 = \sin\theta_1 / \cos\theta_1 = -iv/c$, and thus its coordinates become $p_{\mathsf{S},(\tau_a,x_0)} = (\tau_a, \tan\theta_1 \tau_a)$. The coordinates in $\mathsf{S}^{(d)}$ of $p_{\mathsf{S}^{(d)},(\tau_a^{(d)},x_0^{(d)})}$ are also given by $p_{\mathsf{S}^{(d)},(\tau_a^{(d)},x_0^{(d)})} = (\tau_a^{(d)}, \tan\theta_1 \tau_a^{(d)})$ where $p_{\mathsf{S}^{(d)},(\tau_a^{(d)},x_0^{(d)})}$ is the corresponding point in $\mathsf{S}^{(d)}$ to $p_{\mathsf{S}'^{(d')},(\tau_a'^{(d')},0)} = (\tau_a'^{(d')}, 0)$. Recall that $\tau^{(d)} = \tau$ and $x^{(d)} = x - d$, and that $p_{\mathsf{S}'^{(d')},(\tau_a'^{(d')},0)}$ is represented as $p_{\mathsf{S}',(\tau_a',d')} = (\tau_a', d')$ in $\mathsf{S}'$. The coordinates of $p_{\mathsf{S},(\tau_a,x_\beta)}$, which is the representation in $\mathsf{S}$ of $p_{\mathsf{S}^{(d)},(\tau_a^{(d)},x_0^{(d)})}$, corresponding to $p_{\mathsf{S}',(\tau_a',d')} = (\tau_a', d')$ in $\mathsf{S}'$ are then written as

$$p_{\mathsf{S},(\tau_a,x_\beta)} = (\tau_a, x_\beta) = (\tau_a, d + \tan\theta_1 \tau_a). \tag{32}$$

The relationship between $d'$ and $d$ can be obtained from (27). Eliminating $\tau^{(d)}$ in (27) with $\theta = \theta_1$, we have

$$\sin\theta_1 \tau'^{(d')} + \cos\theta_1 x'^{(d')} = x^{(d)}. \tag{33}$$

If $x^{(d)}$ is fixed at a certain value, say $\alpha$, (33) represents the line equation in $\mathsf{S}'^{(d')}$ corresponding to line $x^{(d)} = \alpha$. Substituting $\tau'^{(d')} = 0$ with $x'^{(d')} = x' - d'$ and $x^{(d)} = x - d$ into (33), it follows that

$$\cos\theta_1 (x' - d') = x - d. \tag{34}$$

Clearly, if $d = 0$, $d' = 0$. Thus $\cos\theta_1 x' = x$, and then

$$\cos\theta_1 d' = d. \tag{35}$$

When $d' = x_a'$, let the point in $\mathsf{S}$ corresponding to $p_{\mathsf{S}',(\tau_a',x_a')}$ be $p_{\mathsf{S},(\tau_a,x_a)}$. From (32),

$$d = x_a - \tan\theta_1 \tau_a. \tag{36}$$

Substituting (36) into (35) yields

$$\cos\theta_1 x_a' = x_a - \tan\theta_1 \tau_a. \tag{37}$$

Then the coordinates of the point $\boldsymbol{p}_{\mathsf{S}',a} = \boldsymbol{p}_{\mathsf{S}',(\tau_a',x_a')}$ in $\mathsf{S}'$ corresponding to $\boldsymbol{p}_{\mathsf{S},a} = \boldsymbol{p}_{\mathsf{S},(\tau_a,x_a)}$ are obtained, from (11), (29), and (37), as

$$\begin{aligned}
\boldsymbol{p}_{S',a} &= \sqrt{1-(v/c)^2}\,[\tau_a,\, x_a + i(v/c)\tau_a]^T \\
&= \frac{1}{\cos\theta_1}[\tau_a,\, x_a - \tan\theta_1\,\tau_a]^T \\
&= \frac{1}{\cos\theta_1}\begin{bmatrix} 1 & 0 \\ -\tan\theta_1 & 1 \end{bmatrix}\boldsymbol{p}_{S,a} \equiv \boldsymbol{T}_{OS}(\theta_1)\boldsymbol{p}_{S,a}.
\end{aligned} \tag{38}$$

Note that in terms of observation points, (38) is valid for every point $(\tau_a, x_a)$, while (10) is valid only for the $\tau'$-axis which is the observation line of $O'$.

In Fig. 5, the crossing $p_{S,cr} = (\tau_a, x_{cr})$ of lines $\tau' = \tau_a'$ and $\tau = \tau_a$ seems to be $(\tau_a, 0)$ in $S$. As a matter of fact, the crossing is below the point $(\tau_a, 0)$. For notational simplicity, $p_{S',(\tau_a',0)}$ and $p_{S,(\tau_a,x_0)}$ are denoted as $p_{S',0}$ and $p_{S,0}$, respectively. The distance $d(p_{S',0})$ and the coordinates of $p_{S,cr}$ can be obtained using Fig. 6, where $p_{S',0}$, $p_{S,0}$, and $p_{S,cr}$ correspond to points 0, 4, and 3, respectively. Note that Fig. 6 employs real, not complex, time axes. The distance $d(p_{S',0})$ and $x_{cr}$ are calculated from the figure as

$$d(p_{S',0}) = \frac{\tau_a}{\cos\theta_1} \tag{39}$$

$$x_{cr} = 2\tau_a \frac{\cos^2\theta_1 - \sqrt{\cos 2\theta_1}}{\sin 2\theta_1}. \tag{40}$$

The derivations are given in Appendix 6.1. Equation (39) is in agreement with (29). When line $\tau = \tau_a$ is rotated in the counter clockwise direction by $\theta_1$ at the center of $p_{S,cr}$, the resultant line is the same as line (31). Hereafter, lines of simultaneity will be approximately plotted as if $x_{cr}$ or $x_{cr}'$ is zero.

From Fig. 7, we can see the physical meaning of distances in the observation system $S\,(=S)$. Fig. 7 shows that an event $e_S(\tau_a, x_a)$ at $p_{S,a} = (\tau_a, x_a)$ is propagating toward the observer $O$. When $\tau = \tau_a$, the event $e_S(\tau_a, x_a)$ occurs at $x = x_a$, and is first observed by the observer $O^{(x_a)}$. Then, it is observed by $O$ when $\tau = \tau_a + ix_a$ if $x_a > 0$ or when $\tau = \tau_a - ix_a$ if $x_a < 0$, i.e., when $d(\tau_a, x_a; \tau, 0) = 0$ with $t > t_a$. Suppose that at present, the time is $\tau = \tau_a$. Then, in Fig. 7, the observer $O$ is at point 0, the coordinates of which are $(\tau_a, 0)$. Consider an event $e_S(\tau, x)$ at a point $(\tau, x)$. If the distance $d(\tau, x; \tau_a, 0)$ for $i\tau \geq i\tau_a$ $(t \leq t_a)$ is imaginary, zero, or real, the event $e_S(\tau, x)$ is a past, present, or future event, respectively, to $O$. If $i\tau < i\tau_a$ $(t > t_a)$, $e_S(\tau, x)$ is a future event to all observers.

*2.3. Time dilation and Length Contradiction*

It is well known that the time (unit) dilation and the length contraction occurs in $S'$ in motion with a constant velocity relative to $S$ being at rest. Using (13) and (14), we have

$$\Delta t' = t_2' - t_1' = \cos\theta_1 \Delta t \tag{41}$$

$$\Delta x' = x_2' - x_1' = \cos\theta_1 \Delta x \tag{42}$$

where $\Delta t = t_2 - t_1$, $\Delta x = x_2 - x_1$, $t_2 > t_1$, $x_2 > x_1$, and $\cos\theta_1$ is given in (11). The same $x$ and the same $t$ were used in the calculation of (41) and (42), respectively. Since $\cos\theta_1 > 1$, thus $\Delta t' > \Delta t$ and $\Delta x' > \Delta x$. Despite the obvious facts by the conventional theory, it is said that the time (unit) dilation and length contraction occur in $S'$. Of course, the inequalities can be reversed according to the relativity principle, which makes us more confused. The time dilation occurs in $S'$ when the time interval is measured as follows [3]. See Fig. 5 (regarding the coordinates as those for $S'$ and $S$, though the coordinates in Fig. 5 are for $\mathsf{S'}$ and $\mathsf{S}$). When $t = t_a$, $O'$ is at position $p_{S',0} = (\tau_a', 0)$ in $S'$ while it appears to be at position $p_{S,0} = (\tau_a, x_0)$ in $S$. Since they met at $\tau = \tau' = 0$, a time $t_a$ passed in $S$ while a time $t_a'$ $(=\sqrt{1-(v/c)^2}\, t_a < t_a)$ passed in $S'$. It shows time dilation in $S'$. Note that $S'$ and $S$ are coordinate systems, not observation systems, and only the points on the $\tau$- and $\tau'$-axes are associated with the observations of $O$ and $O'$, respectively.

The facts that $\Delta t' > \Delta t$ and that $\Delta t' < \Delta t$ contradict each other. The reason for the contradiction can be clearly seen from Fig. 8. The coordinates of points 1 and 2 in $S$ are $(\tau_1, x_a)$ and $(\tau_2, x_a)$. Let the coordinates in $S'$ corresponding to points 1 and 2 be $(\tau_1', x_{a1}')$ and $(\tau_2', x_{a2}')$. Since distances are invariant, the distance from point 1 to point 2 is equal to that from $(\tau_1', x_{a1}')$ to $(\tau_2', x_{a2}')$ and so

$$\begin{aligned} d(\tau_2, x_a; \tau_1, x_a) &= d(\tau_2', x_{a2}'; \tau_1', x_{a1}') \\ &= \sqrt{(\tau_2' - \tau_1')^2 + (x_{a2}' - x_{a1}')^2} \end{aligned} \tag{43}$$

Since $(x_{a2}' - x_{a1}')^2 > 0$, $|\tau_2' - \tau_1'| > |\tau_2 - \tau_1|$. Thus, $\Delta t' (= t_2' - t_1') > \Delta t (= t_2 - t_1)$. It can be similarly shown that $\Delta t (= t_4 - t_3) > \Delta t' (= t_4' - t_3')$ from points 3 and 4 in $S'$ where the coordinates of points 3 and 4 are $(\tau_3', x_a')$, and $(\tau_4', x_a')$, and the coordinates in $S$ corresponding to the points are $(\tau_3, x_{a1})$, and $(\tau_4, x_{a2})$. The space and time have the duality, as mentioned in Section 2.1. Hence, the length contraction can be similarly explained.

The contradiction in the conventional theory results from no distinction between the coordinate system and the observation system. It is straightforward to see the time dilation and the length contraction from (38). The time dilation and the length contraction occur in $S'$ according to (38). There are no contradictions when observing time intervals and lengths in the observation systems $S'$ and $S$.

*2.4. Special Relativity in the Real Number Space*

In this Section, the coordinates of points will be represented in the real number space as in the conventional approach. Distances are invariant with respect to the rotation between the coordinate systems $S'$ and $S$, and thus

$$\|\boldsymbol{p}_{S',a}\| = \|\boldsymbol{p}_{S,a}\| = \sqrt{-c^2 t_a^2 + x_a^2}, \qquad (44)$$

which shows that if $\|\boldsymbol{p}_{S,a}\|$ is imaginary, $\|\boldsymbol{p}_{S',a}\|$ is also imaginary. From (44), we have

$$\begin{aligned} l(\boldsymbol{p}_{S',a}) + x_a^2 &= c^2 t_a^2, &\text{for } c^2 t_a^2 \geq x_a^2 \\ l(\boldsymbol{p}_{S',a}) + c^2 t_a^2 &= x_a^2, &\text{for } c^2 t_a^2 \leq x_a^2 \end{aligned} \qquad (45)$$

Equation (45) indicates that the three sides of lengths $l(p_{S,a})$, $l(x_a)$, and $l(ct_a)$ form a right-angled triangle with the hypotenuse of length $l(ct_a)$ when $l(ct_a) > l(x_a)$ where $l(\beta)$ is the absolute value of a number $\beta$, i.e., $l(\beta) = |\beta|$. Referring to Fig. 1, we define $\tilde{\theta}_1$ such that for $l(ct_1) \geq l(x_1)$, $\cos\tilde{\theta}_1$ and $\sin\tilde{\theta}_1$ are given by

$$\cos\tilde{\theta}_1 = \frac{d(\boldsymbol{p}_{S,1})}{ict_1} = \sqrt{1-(v/c)^2} \qquad (46)$$

$$\sin\tilde{\theta}_1 = \frac{x_1}{ct_1} = \frac{v}{c}. \qquad (47)$$

Equations (13) and (14) can be expressed using (46) and (47) as

$$\cos\tilde{\theta}_1 \, ct_a' = ct_a - \sin\tilde{\theta}_1 \, x_a \qquad (48)$$

$$\cos\tilde{\theta}_1 \, x_a' = -\sin\tilde{\theta}_1 \, ct_a + x_a \, . \qquad (49)$$

Let us normalize space and time variables and quantities in $S$ with respect to $c$ in such a way that $\bar{t} = ct$, $\bar{v} = v/c$, $\bar{x}_a = x_a$, and so on. Similar normalizations are also made in $S'$. Using the normalizations, for $\cos\tilde{\theta}_1 \neq 0$, (48) and (49) can be expressed in vector form as

$$\overline{\boldsymbol{p}}_{S',a} = \boldsymbol{T}_{RS}(\tilde{\theta}_1)\overline{\boldsymbol{p}}_{S,a} \qquad (50)$$

where

$$\overline{\boldsymbol{p}}_{S',a} = [\bar{t}_a', \bar{x}_a']^T \qquad (51)$$

$$\bar{p}_{S,a} = [\bar{t}_a, \bar{x}_a]^T \tag{52}$$

$$T_{RS}(\tilde{\theta}_1) = \frac{1}{\cos\tilde{\theta}_1}\begin{bmatrix} 1 & -\sin\tilde{\theta}_1 \\ -\sin\tilde{\theta}_1 & 1 \end{bmatrix}. \tag{53}$$

It is straightforward to see that $T_{RS}(-\tilde{\theta}_1)T_{RS}(\tilde{\theta}_1) = I$, which implies that the transformation (50) satisfies the relativity. Note that $T_{RS}(\tilde{\theta}_1)$ is a symmetric matrix, i.e., $T_{RS}^T(\tilde{\theta}_1) = T_{RS}(\tilde{\theta}_1)$. Equation (50) also shows the duality between space and time, being invariant with respect to the interchange of the space and time variables. Using hyperbolic functions, (50) is rewritten as [2]

$$\bar{p}_{S',a} = T_{HF}(\tilde{\theta}_2)\bar{p}_{S,a} \tag{54}$$

where

$$T_{HF}(\tilde{\theta}_2) = \begin{bmatrix} \cosh\tilde{\theta}_2 & -\sinh\tilde{\theta}_2 \\ -\sinh\tilde{\theta}_2 & \cosh\tilde{\theta}_2 \end{bmatrix} \tag{55}$$

with $\cosh\tilde{\theta}_2 = [1-(v/c)^2]^{-1/2}$ and $\sinh\tilde{\theta}_2 = (v/c)[1-(v/c)^2]^{-1/2}$. Given $p_{S,a}$, the coordinate values of $p_{S',a}$ can be found using (48) and (49). For simplicity, we assume that $t_a, x_a \geq 0$ and $0 \leq \tilde{\theta}_1 < 90°$.

The coordinate values of $p_{S',a}$ are shown in Fig. 9 where $a$ denotes $\tilde{\theta}_1$. Two axes, $\mathrm{val}(ct')$ and $\mathrm{val}(x')$, are introduced to indicate the $ct'$- and $x'$-values. In Fig. 9(a), which illustrates the $ct'$-coordinate value, $d(p_{S,a},1) = d(2,1) = x_a$, and the quantities $\sin\tilde{\theta}_1 x_a$ and $\cos\tilde{\theta}_1 ct_a'$ ($= ct_a - \sin\tilde{\theta}_1 x_a$) are equal to $d(1,3)$ and $d(3)$, respectively, where $d(\alpha,\beta)$ denotes the distance from point $\beta$ to point $\alpha$, and the origin is omitted, for example $d(3) = d(3,O)$. Then $ct_a'$ is given by $ct_a' = d(4)$. The value of $x_a'$ is indicated on the axis $\mathrm{val}(x')$ in Fig. 9(b). In the figure, $d(p_{S,a},5) = d(6,5) = ct_a$. The quantities $\sin\tilde{\theta}_1 ct_a$ and $\cos\tilde{\theta}_1 x_a'$ ($= x_a - \sin\tilde{\theta}_1 ct_a$) are equal to $d(5,7)$ and $d(7)$, respectively, and $x_a'$ is then given by $x_a' = d(8)$.

Consider a case that time in $S$ is fixed at $t = t_a$ and $x$ varies. Changing $x_a$, $x_a'$ and $t_a'$ in (48) and (49) into $x$, $x'$ and $t'$ and eliminating $x$, we have

$$ct' + \sin\tilde{\theta}_1 x' = \cos\tilde{\theta}_1 ct_a. \tag{56}$$

The $ct'$-coordinate when $x' = 0$ is denoted as $ct_a'$, which is, from (56), related to $ct_a$ by $ct_a' = \cos\tilde{\theta}_1 ct_a$. When $x$ is fixed at $x = x_a$, points in $S'$ corresponding to line $x = x_a$ are computed as

$$x' + \sin \tilde{\theta}_1 \, ct' = \cos \tilde{\theta}_1 \, x_a. \tag{57}$$

Note that (56) and (57) represent just the coordinates in $S'$, not observation points of $O'$. Equations (56) and (57) represent line equations. Lines (56) and (57) are illustrated in Fig. 10, where the axis val$(x')$ overlaps the $x$-axis, and in Fig. 11, where the axis val$(ct')$ overlaps the $ct$-axis, respectively. The lines (56) and (57) perpendicularly intersect the axis val$(ct')$ in Fig. 10 and the axis val$(x')$ in Fig. 11, respectively. Let us explain Fig. 10 of line (56). The distance $d(2,3) = x'$, $d(1,3) = \sin \tilde{\theta}_1 \, x'$, and $d(3) = ct'$. Thus, $ct' + \sin \tilde{\theta}_1 \, x' = d(3) + d(1,3) = ct_a' = \cos \tilde{\theta}_1 \, ct_a$. As shown in Fig. 10, the simultaneous events at time $t = t_a$ in $S$ appear to be past $(t' < t_a')$, present $(t' = t_a')$, and future $(t' > t_a')$ events in $S'$. It is very unnatural, though the conventional theory explains that the unnaturalness is due to the relativity of simultaneity. One of the two coordinates systems $S$ and $S'$ is not equal to its observation system.

## 2.5. Coordinate Bases and the True Space-Time Space

A point $\boldsymbol{p}_{S,a} = [\tau_a, x_a]^T$ is written as

$$\begin{bmatrix} \tau_a \\ x_a \end{bmatrix} = \begin{bmatrix} 1 & 0 \\ 0 & 1 \end{bmatrix} \begin{bmatrix} \tau_a \\ x_a \end{bmatrix}$$
$$= \boldsymbol{U}_S \begin{bmatrix} \tau_a \\ x_a \end{bmatrix} = \boldsymbol{u}_\tau \tau_a + \boldsymbol{u}_x x_a \tag{58}$$

where

$$\boldsymbol{U}_S = [\boldsymbol{u}_\tau, \boldsymbol{u}_x] \tag{59}$$

with $\boldsymbol{u}_\tau = [1, 0]^T$ and $\boldsymbol{u}_x = [0, 1]^T$. The columns of $\boldsymbol{U}_S$ form the coordinate basis of $S$ and $\boldsymbol{p}_{S,a} = [\tau_a, x_a]^T$ is the coordinate vector with respect to the ordered basis vectors $\boldsymbol{u}_\tau$ and $\boldsymbol{u}_x$ [8]. The coordinate vector $\boldsymbol{p}_{S',a}$ in $S'$ is given as (10). In Fig. 1, the coordinate system $S'$ is rotated by $\theta_1$ in the counter clockwise direction with respect to $S$ and then the basis of $S'$ relative to $S$ is given by

$$\boldsymbol{U}_{S'}(\theta_1) = \begin{bmatrix} \cos \theta_1 & -\sin \theta_1 \\ \sin \theta_1 & \cos \theta_1 \end{bmatrix}. \tag{60}$$

Every coordinate system that belongs to an inertial system, say an inertial system $S'$, employs the same basis, $\boldsymbol{U}_{S'}(\theta_1)$. In other words

$$\boldsymbol{U}_{S'}(\theta_1) = \boldsymbol{U}_{S'^{(d')}}(\theta_1) = \boldsymbol{U}_{S'}(\theta_1). \tag{61}$$

Let the vector $\boldsymbol{p}_{S',a}$ be the coordinate vector in $S'$ of point $p_{S,a}$. The representation in $S$ of $\boldsymbol{p}_{S',a}$ is written as

$$\boldsymbol{U}_{S'}(\theta_1)\boldsymbol{p}_{S',a} = \boldsymbol{U}_{S'}(\theta_1)\boldsymbol{T}(\theta_1)\boldsymbol{p}_{S,a} = \boldsymbol{p}_{S,a}. \tag{62}$$

If $S'$ has the identity basis $\boldsymbol{I}$, the basis of $S$ becomes

$$\boldsymbol{U}_S(-\theta_1) = \begin{bmatrix} \cos\theta_1 & \sin\theta_1 \\ -\sin\theta_1 & \cos\theta_1 \end{bmatrix} \tag{63}$$

and the transformation matrix form $S'$ to $S$ becomes $\boldsymbol{T}(-\theta_1)$. It is easy to see that $\boldsymbol{U}_S(-\theta_1)\boldsymbol{T}(-\theta_1) = \boldsymbol{I}$. Then,

$$\boldsymbol{U}_S(-\theta_1)\boldsymbol{p}_{S,a} = \boldsymbol{U}_S(-\theta_1)\boldsymbol{T}(-\theta_1)\boldsymbol{p}_{S',a} = \boldsymbol{p}_{S',a}, \tag{64}$$

which also shows the invariance. Though the representation of a point is different depending on a coordinate system applied, the point itself is invariant in the sense of the equalities (61) and (64).

The time and the length scales in $S'$ should be defined. Two conditions are required to define both the scales. We already have one condition by saying that the velocity of $S$ seen from $S'$ is $-v$. The light speed is assumed to be $c$ in $S'$. As a matter of fact, the assumption can be regarded as another condition to define the scales. These settings are necessary such that both $S$ and $S'$ have the same physical conditions. The principle of relativity can hold under the same conditions. Once the time and the length scales have been set up, no more conditions are required to set up the time or the length scale in $S'$. In other words, the clock and the ruler for $S'$ have been made, in accordance with both the conditions.

Within only the level of the coordinate systems, $S$ and $S'$ appear to be equivalent in relativity. If seeing $S$ and $S'$ in connection with the observation systems, however, we can identify differences between $S$ and $S'$. If neither $S$ nor $S'$ is equal to its observation system, (15) does not have any physical meaning. One coordinate system must be identical to its observation system. In terms of operation, the velocity of light becomes a constant $c$ in any direction if Einstein synchronization is employed under the axiom of the constant round-trip velocity of light [9]. Carrying out Einstein synchronization in an inertial system, we can obtain the same coordinate system as its observation system in which the light speed appears to be constant in every direction irrespective of its actual one-way speeds. As far as the transformation (15) is concerned, however, only one coordinate system is identical to its observation system, the other coordinate system being dependent on it. The dependent coordinate system is different from its observation system. For example, when $S = \mathcal{S}$, the point $p_{S',a} = (\tau_a', x_a')$ in $S'$ corresponding to $p_{S,a} = (\tau_a, x_a)$, which is given by (10), is generally different from that in $\mathcal{S}'$, which is given by (38).

When introducing observation systems, the equality in relativity between $S$ and $S'$, on which the conventional theory of special relativity is based, no longer holds, and one coordinate system becomes dependent on the other which is equal to its observation system. Equation (38) is derived under the assumption that $S = \mathcal{S}$. Revisiting Fig. 5, we see that the set $\tau' = \tau_a'$ of simultaneous points in $\mathcal{S}'$ is represented as line $\tau = \tau_a = \tau_{a1}$ in $\mathcal{S}$. Then

$$|\tau_a'| < |\tau_{a1}|. \tag{65}$$

As mentioned in Section 2.2, the accurate $x$-coordinate of the crossing of lines $\tau = \tau_{a1}$ and $\tau' = \tau_a'$ is not zero, and can be obtained from (40). On the contrary, if $S' = \mathcal{S}'$, the relationship (38) is reversed as

$$\boldsymbol{p}_{S,a} = \boldsymbol{T}_{OS}(-\theta_1)\boldsymbol{p}_{S',a} \tag{66}$$

and then the time dilation and the length contraction occur in $S$. The set of observations points in $\mathcal{S}$ corresponding to line $\tau' = \tau_a'$ in $\mathcal{S}'$ is no longer line $\tau = \tau_a = \tau_{a1}$, and becomes line $\tau = \tau_{a2}$, as shown in Fig. 12, where $\cos(-\theta_1)\tau_{a2} = \cos\theta_1 \tau_{a2} = \tau_a'$. Then

$$|\tau_a'| > |\tau_{a2}|. \tag{67}$$

When $S' = \mathcal{S}'$, the $x'$-coordinate of the crossing of lines $\tau = \tau_{a2}$ and $\tau' = \tau_a'$ is not zero, and can be obtained from (40) with the change of $x_{cr}$, $ct_a$, and $\theta_1$ into $x_{cr}'$, $ct_a'$, and $-\theta_1$, respectively. Equations (65) and (67) contradict each other, showing that the coordinates depend on the selection of the observation system. The equality in terms of the transformation between $S$ and $S'$ does not hold when considering it in connection with observations systems.

The conventional theory of special relativity deals with the transformation from $S$ to $S'$ based on the two axioms, under the condition that $S = \mathcal{S}$. Coordinates obtained according to the conventional theory are the representations in $S'$ of points in $S$. The coordinate system $S'$ is different from $\mathcal{S}'$, as mentioned above. The misunderstanding that $\mathcal{S}' = S'$ has led to the relativity of simultaneity. The equality in relativity, together with the relativity of simultaneity, causes contradictions and paradoxes. The perception that $\mathcal{S}' = S'$ has also led to the equality in relativity. The physical laws can be represented as the same forms in every coordinate system, but inertial systems are not equivalent.

One observation system which has no time dilation with respect to the other is the reference system, from which the coordinate system of an observer in the latter inertial system is determined. However, when introducing the operation of Einstein synchronization, unfortunately, we cannot know which system, $\mathcal{S}$ or $\mathcal{S}'$, is the reference system by comparing time intervals in $\mathcal{S}$ and $\mathcal{S}'$ because to choose any one coordinate system to be the same as its observation system always results in the time dilation in the other system. We will take a space trip in Section 3 in search of the true space and time not relying on any conventions of synchronization. Here, we just define the reference system as the one that has a larger time interval than the other, although it depends on the choice. The reference system has the identity basis. In the true space-time space, which can be regarded as being at rest relative to every inertial system, its observation system is the reference system with respect to every inertial system, being identical to the coordinate system of an observer, and the coordinate basis is the identity relative to any inertial system. Time in the true space runs faster than in any other inertial system.

It is seen from (38) that the light speed in $S'$ is given by

$$c_+' = c - v \tag{68}$$

$$c_-' = c + v \tag{69}$$

where $c_+'$ and $c_-'$ are the light speeds in the $+x'$-direction and $-x'$-direction, respectively. Though the two-one-way average speed of light in $S'$, which is defined as $(c_+' + c_-')/2$, is $c$, its round-trip average speed is different from $c$. It is obvious that when $\tau = \tau' = 0$, not only $O'$ met $O$, but also $O'^{(d')}$ met $O^{(d)}$, which means that $e_{S'^{(d')}}(0,0) = e_{S^{(d)}}(0,0)$. Based on the obvious fact, (38) was derived. If the round-trip speed of light has been clearly proven to be constant in every inertial system, the constancy of the round-trip speed can hold in $S'$ by simply changing the scale of $x'$-coordinate in (38) (with subscript 'a' omitted) such that

$$\begin{bmatrix} \tau' \\ x' \end{bmatrix} = \frac{1}{\cos\theta_1}\begin{bmatrix} 1 & 0 \\ -\cos^2\theta_1 \tan\theta_1 & \cos^2\theta_1 \end{bmatrix}\begin{bmatrix} \tau \\ x \end{bmatrix}$$
$$= \begin{bmatrix} \cos^{-1}\theta_1 & 0 \\ -\sin\theta_1 & \cos\theta \end{bmatrix}\begin{bmatrix} \tau \\ x \end{bmatrix}. \tag{70}$$

Then the relationship between $x'$ and $x$ is reduced to that in (15). The same transformation formulations as (70) are found in [9–12]. The velocity of $S'$ seen from $S$ is still $v$. However, the velocity of $S$ seen from $S'$ is no longer $-v$ due to the scale change and is changed to $-\cos^2\theta_1 v$.

Someone may say that the constancy of the light speed is experimentally verified. The light speed constancy is well explained in [9]. The reason for the constancy is that the scales of time and space in an inertial system, say $S'$, are set such that the light speed appears to be constant in every direction. In other words, the scales of time and space are determined by the light speed constancy. In an inertial system $S'$, the space and time scales are set such that the round trip speed appears constant, and time is set according to Einstein synchronization. Then, the light speed appears constant in every direction, as explained in [9], and the observation system $S'$ appears to be the same as the coordinate system $S'$. Fig. 3 shows how time and space are set in order for the light speed to be invariant. A line in $S$ of (21) with a slope of $\eta$ is rotated so that it appears to be the line in $S'$ of (22) with the slope given by (23). If light speeds are measured as a constant in every inertial system, we, human being, may be created so that we see space and time, regardless of the movement with a constant velocity, as if the light speed is constant.

Events occur simultaneously in every inertial system, as shown in Section 2.2. It can be readily shown, in another way, using Fig. 13. An observer $O'^{(d')}$ in an inertial system $S'$, who is at $x = d$ and $t = 0$, moves to $x = vt_1 + d$ at $t = t_1$. Every observer in the inertial system $S'$ is on the line $t = t_1$ in the inertial system $S$, which is an obvious fact irrespective of light speeds because it represents only the relative movement of the inertial system $S'$ with a velocity of $v$. Of course, the velocity $v$ is the one observed by the inertial system $S$. Clearly, if the events at $t = t_1$ in the inertial system $S$ occur actually simultaneously, they occur actually simultaneously in the inertial system $S'$ as well, though the latter system does not see them as simultaneous events according to the space and time setting for the light speed constancy as shown in Fig. 3. The existence of the true space and time is independent of our inability of not being able to experimentally find them with the present knowledge and techniques. It is believed that the true space and time exist. Only in the true space and time, events occur actually simultaneously. When applying the Einstein's two axioms, the transformation relationship between the coordinate systems $S$ and $S'$ becomes (15). Accordingly simultaneous events become different from one coordinate system to another (see Figs. 3 and 10). According to the principle of relativity, the conventional theory takes as actual simultaneous events each set of events that appear simultaneous in each coordinate system, which is described as the relativity of simultaneity. It is the reason why the conventional theory has some paradoxes concerning simultaneity. If neither of the inertial systems $S$ and $S'$ is the true space and time, the events that one or the other system sees as being simultaneous, under the light speed constancy, are actually not simultaneous. We, inertial systems, may not be able to see true simultaneous events simultaneously before we find the true space and time.

Consider the case that an observer $O'$ is moving along the $x$-axis with a velocity of $v = c$ relative to an observer $O$ (refer to Fig. 1). As $v$ approaches $c$, $\cos\theta_1$ in (11) tends to infinity. It is shown in Appendix 6.2 that when $v = c$, the points in $S'$ which represent points in $S$ are only the origin of $S'$. Since $S$ and $S'$ cannot represent each other's points except the origin in their coordinate systems, they are essentially independent in terms of the representation of points. Hence, two inertial systems in relative motion with a velocity of $c$ are independent of each other in the representation of points.

Now we know what the true space-time space is. However, we do not know where it is. We have lost the true time and space since the appearance of the special relativity. However, we believe that the true space-time space will exist. A space trip to investigate the universe from its beginning to the present, together with the belief in the existence, the new approach, and just the special relativity that has made the true space and time disappear, may lead us to find a clue for the true space and time.

## 3. The Universe and Time

Let us start the space trip in search of the true space and time. Unless the true space and time exist, we do not have to take a trip and have to live in a relative world without the true space and time. It is assumed that the true space and time exist. Our first destination is the beginning of the universe.

Let $t_u$ and $O_u$ denote the true (or universe) time and the origin of the universe, respectively. Suppose that two objects, $O_1$ and $O_2$, meet at the universe origin, $O_u = O_1 = O_2$, when $t_u = t_{u1} = t_{u2} = 0$. One object $O_2$ is moving with a constant velocity of $v$ with respect to the other $O_1$. Unless $v = c$, as shown in Sections 2.2 and 2.5, the position of an observation point in $S_1$ is different from that in $S_2$ where $S_1$ and $S_2$ are the observation systems of the inertial systems $S_1$ and $S_2$, respectively. Thus, the velocity that can allow the true time and space to exist is only $v = c$. Consider all the objects that are moving in every radial direction with a relative speed of $c$ with respect to $O_u$. At $t_u = t_{u,p}$, they all reach the surface of a sphere of radius $r = ct_{u,p}$. Since $r = ct_{u,p}$, a point $(x, y, z)$ on the surface of the sphere in the rectangular coordinate system with the origin $O_u$ has the relationship

$$\sqrt{x^2 + y^2 + z^2} = ct_{u,p}. \tag{71}$$

The object on a point $(x, y, z)$ of the sphere sees an event $e[\tau_{u,p}, (x, y, z)]$ in its own coordinate system where $\tau_{u,p} = ict_{u,p}$. The coordinate systems of the objects are independent of each others, as explained in Section 2.5. Hence, the set $S_u$ of the observation points for all the object is the true space-time space. The origin of the true space-time space is $O_u$. However, any objects cannot see the true space-time space since each others' observation points reduce to zero, as shown in Appendix 6.2. However, we can see the whole universe. In fact, each object virtually moves along its own time axis, not a space axis. For example, the direction in which an object $O_1$ is moving is the same as its time axis, the $ct_{u1}$-axis; see Fig. 14 where the time and space axes are represented as dotted lines and solid lines, respectively. Hereafter, in the term 'space-time space', we call the first 'space' a 'pure space' to distinguish it from the second 'space', if necessary. Time becomes pure space at the instant that it becomes the present. Revisit Fig. 7. An observer $O^{(d)}$ in Fig. 7, who is at $x = d$, does not move in the pure space, but seems to move with a velocity of $c$ in the space-time space as time passes. In other words, when $\tau = 0$, $O^{(d)}$ is at $\tau = 0$, and when $\tau = \tau_a$, it is moved to $\tau = \tau_a$. In the space-time space, we can regard the pure space as a set of points where events occur and occurred, and the time space as a set of points before the occurrence. Before $t = 0$, the space-time space of $0 \geq i\tau \geq i\tau_a$, i.e., $0 \leq t \leq t_a$, was the time space. But it is changed into the pure space at the instant $\tau = \tau_a$. All events occur in the pure space, not in the time space, in the sense that time is changed into pure space at the instant of occurrence. The set of line $\tau = \tau_a$ is changed into pure space at the instant that the simultaneous events $e(\tau_a, x)$ appear at $\tau = \tau_a$, and then the events are propagated in the pure space. In fact, in Fig. 14, $O_1$ is moving along the $ct_{u1}$-axis. Since time becomes pure space at the instant that it becomes the present time, we can equivalently consider that $O_1$ is not moving and just time passed. In consequence, $O_1$ can observe the true space (and time space) with the origin $O_u$, where the phrase including 'time' is bracketed since time is changed into pure space at the instant of occurrence.

Surprisingly, (71) implies the expansion of the universe (see Fig. 14). The universe is expanding with an increase in $t_u$. The expansion of the universe is equivalent to the increase of the universe time. If seeing it from another viewpoint, the universe time is being created by the universe expansion, and is the universal event that things in the universe occur simultaneously. In fact, time is equivalent to change. If anything is not changing, there is no time. Change creates time. The change in the universe is the expansion of the universe so that time flows in the whole universe. Change is made in the space, and so time cannot exist alone without space. The true natures of time and space are the same. Only their appearances are different. The appearance of time is imaginary if the appearance of space is real. Moreover, in terms of the special relativity, the space and time have the duality in that $\tau$ and $ix$ can be interchanged, as explained in Section 2.1. It is the universe expansion speed that is the light speed in the true space. In Fig. 1, we set $\tau = ict$, considering that electromagnetic waves propagate with a light speed. In fact, the $c$ is the universe expansion speed if the $\tau$-axis is connected to $O_u$. The universe itself expanding in every radial direction is the true space. The whole true space is moving with the light speed so that events are changing and spreading with the same speed. Though the special relativity has made us lose the true space and time, it now not only helps us to find the true space, but also provides surprising information.

The Hubble constant [6] can be derived using the expanding universe model. At certain $t_{u,p}$, two objects are located at $\boldsymbol{p}_1$ and $\boldsymbol{p}_2$, and the distance between them is $d$, as shown in Fig. 15. The angle between $\boldsymbol{p}_1$ and $\boldsymbol{p}_2$ is $\alpha$ and $\|\boldsymbol{p}_1\| = \|\boldsymbol{p}_2\| = ct_{u,p}$. It is seen from Fig. 15 that after $\Delta t_u$, an increase $\Delta d$ in the distance is

$$\Delta d = \frac{c \Delta t_u}{c t_{u,p}} d \tag{72}$$

and thus

$$v = \frac{\Delta d}{\Delta t_u} = \frac{1}{t_{u,p}} d = Hd \tag{73}$$

where the Hubble constant is given by

$$H = \frac{1}{t_{u,p}}. \tag{74}$$

It is well known [6] that the moving-away speed between two galaxies is proportional to the distance between them and that the reciprocal of the Hubble constant for the galaxies that are located near the boundary of the universe is equal to the age of the universe. Equations (73) and (74) are consistent with these observations. We can also know the size of the universe if the value of the Hubble constant is known.

## 4. Resolution of Twin Paradox

We returned to the Earth by an instantaneous movement after the space trip. We wonder if we are older or younger than our twin sisters and brothers. The twin paradox [2, 5] can be described as follows, based on the previous understanding of special relativity. There is a pair of twins on the Earth. One of them, $A'$, takes a space trip to a distant star with a speed close to the light speed, while the other, $A$, stays at home. A clock in motion relative to an inertial observer runs slower than the clock of the observer at rest. According to it, when $A'$ returns after the trip, $A'$ will be younger than $A$. However, the motion is relative to each other. When considering $A'$ to be at rest, the age of $A$ should be less than that of $A'$.

The twin paradox can be easily seen in the observation systems, $S'$ for $A'$ and $S$ for $A$. Fig. 16 illustrates examples of paradox. Suppose that $A'$ moves from point 1' in $S'$ to $S$ by instantaneous movement. The observer $A'$ regards $S$ as moving and $S'$ as being at rest. Then, $A'$ is located at point 1 in $S$ and becomes older than $A$. On the other hand, $A$ regards $S'$ as moving and $S$ as being at rest. If $A$ moves from point 1 to $S'$, $A$ is at point 2' and becomes older than $A'$. It is a paradox. The twin paradox can be seen in another way. If $A'$ continuously moves between the two systems so that $A'$ is at points, in order, 1', 1, 2', 2, 3', and so on, $A'$ goes into the past whenever $A'$ moves. We cannot go into the past.

As shown in Fig. 17, the events at $\tau' = \tau_a'$ in $S'$ appear to be the events at $\tau = \tau_{a1} = \tau_a$ according to the viewpoint of $S$ that regards itself as being at rest where $\cos\theta_1 \tau_a' = \tau_a$. On the contrary, according to the viewpoint of $S'$ that regards itself as being at rest, the events at $\tau' = \tau_a'$ in $S'$ appear to be the events at $\tau = \tau_{a2}$ in $S$ where $\cos(-\theta_1)\tau_{a2} = \tau_a'$. The points in $S$ corresponding to line $\tau' = \tau_a'$ are different according to who is at rest relative to the other. The question is which place, point 1 or 2 in Fig. 17, $A'$ in $S'$ goes to when at $\tau' = \tau_a'$ returning to the Earth by instantaneous movement. We, $A'$, think that we will returns to point 2, while our twins on the Earth think that we will return to point 1. The true space we have found in Section 3 tells us which one is right. The motion of the Earth with respect to the true space is assumed to be negligible as compared with the size and the expansion speed of the universe. Then, the $\tau$-axis of the Earth can be considered to be connected to the origin $O_u$ of the universe when going back to the past. The space that the Earth sees is then the true space. The line $\tau' = \tau_a'$ in the spaceship corresponds to the line $\tau = \tau_{a1}$ on the Earth. Therefore, we return to point 1, not point 2. The twin paradox has been resolved.

We returned to the Earth, the true space-time space. We are younger than our twins. We now live in the true space.

## 5. Conclusions

The conventional theory of special relativity could not have clearly resolved the contradictions and the paradoxes involving two or more observers, which results from no knowledge on observation systems or no distinction between the coordinate system, which belongs to a specific observer, and the observation system, which includes the observation points of all observers in the same inertial system. The new approach has solved the problems associated with the conventional theory without any contradiction, introducing coordinate systems with imaginary time axes, observation systems, and coordinate bases. In the new coordinate system which lies in CES, relative motion is simply described by the rotation of one coordinate system with respect to the other. The transformation from one coordinate system to another is easily obtained with a complex angle which is dependent on the relative velocity. In observation systems, the absolute simultaneity is shown which means that events occurs simultaneously in every inertial system. Time in the reference inertial system runs faster than in the other. The true space-time space is the reference inertial system in which its observation system is the same as the coordinate system and time runs faster than in any other inertial system.

When applying the new approach to the universe under the assumption of the existence of the true space and time, it shows that the universe is expanding, creating the universe time. Moreover, the light speed is the expansion speed of the universe. The universe itself expanding in every radial direction with the light speed is the true space-time space. Time is changed into space at the instant of occurrence of events. The true natures of time and space are the same. Only their appearances are different, time being imaginary while space is real.

# 6. Appendix

*6.1. Derivations of (39) and (40)*

Fig. 6 is used to obtain $d(p_{S',0})$ and $x_{cr}$, which in the figure correspond to $id(0)$ and the $x$-coordinate of point 3, respectively. Note that in Fig. 6, the time axes of the coordinate systems of the observers $O$ and $O'$ are real. The $ct'$- and $ct$-coordinates of points 0 and 2 are $ct_a'$ and $ct_a$, respectively. Distance $d(0)$ equals $d(1)$, $d(1,2) = d(4,2) = vt_a$, and the angle between segments $(1, O)$ and $(1,2)$ is a right angle. The quantity $v/c$ is expressed as

$$v/c = i\sin\theta_1 / \cos\theta_1 = i\tan\theta_1. \tag{75}$$

The angles $\alpha$ and $\beta$ are related to $\theta_1$ by

$$\cos\alpha = \sqrt{1-(v/c)^2} = 1/\cos\theta_1 \tag{76}$$

$$\cos\beta = 1/\sqrt{1+(v/c)^2} = \cos\theta_1 / \sqrt{\cos 2\theta_1} \tag{77}$$

$$\sin\beta = (v/c)/\sqrt{1+(v/c)^2} = i\sin\theta_1 / \sqrt{\cos 2\theta_1}. \tag{78}$$

1) Derivation of (39)

Since $d(p_{S',0}) = id(0) = id(1)$, $d(p_{S',0}) = id(2)\cos\alpha = ict_a / \cos\theta_1 = \tau_a / \cos\theta_1$.

2) Derivation of (40)

The coordinates in the $(ct, x)$-system of point 0 are

$$p_0 = (ct_0, x_0) = (ct_a \cos\alpha\cos\beta, ct_a \cos\alpha\sin\beta). \tag{79}$$

The line that crosses point 0 and is perpendicular to the $ct'$-axis is expressed as

$$x = -\frac{\cos\beta}{\sin\beta}(ct - ct_a \cos\alpha\cos\beta) + ct_a \cos\alpha\sin\beta. \tag{80}$$

When $ct = ct_a$, the $x$ becomes

$$x_{cr} = \frac{\cos\alpha - \cos\beta}{\sin\beta} ct_a. \tag{81}$$

Since $\alpha > \beta$, $x_{cr} < 0$. Substituting (76)–(78) into (79) with the use of $\tau_a = ict_a$ yields (40).

*6.2. Case of $v = c$*

Eigendecomposing the transformation matrix $\boldsymbol{T}(\theta)$, which is given in (16), we have

$$\boldsymbol{T}(\theta)\boldsymbol{E} = \boldsymbol{E}\boldsymbol{\Lambda}_\theta \tag{82}$$

where

$$E = [e_1, e_2] = \frac{1}{\sqrt{2}} \begin{bmatrix} 1 & 1 \\ i & -i \end{bmatrix} \tag{83}$$

$$\Lambda_\theta = \begin{bmatrix} \lambda_1 & 0 \\ 0 & \lambda_2 \end{bmatrix} = \begin{bmatrix} \cos\theta + i\sin\theta & 0 \\ 0 & \cos\theta - i\sin\theta \end{bmatrix}. \tag{84}$$

It is easy to see that

$$E^H E = I \tag{85}$$

where $H$ designates complex conjugate transpose. Equation (85) indicates that the eigenvector matrix $E$ is unitary. From (82) and (85),

$$T(\theta) = E\Lambda_\theta E^H. \tag{86}$$

Substituting (86) with $\theta = \theta_1$ into (10) gives

$$p_{S',a} = E\Lambda_{\theta_1} E^H p_{S,a}. \tag{87}$$

Multiplying (87) by $E^H$, we have

$$E^H p_{S',a} = \Lambda_{\theta_1} E^H p_{S,a}. \tag{88}$$

Equation (88) is rewritten as

$$\begin{bmatrix} e_1^H p_{S',a} \\ e_2^H p_{S',a} \end{bmatrix} = \begin{bmatrix} \lambda_1 e_1^H p_{S,a} \\ \lambda_2 e_2^H p_{S,a} \end{bmatrix}. \tag{89}$$

As $v$ approaches $c$, from (83), (84), (11), and (12),

$$\lim_{\cos\theta_1 \to \infty} \lambda_2 e_2^H p_{S,a} = \frac{1}{\sqrt{2}} \lim_{v \to c} \frac{1 - v/c}{\sqrt{1-(v/c)^2}} e_2^H p_{S,a} = 0. \tag{90}$$

When $v$ tends to $c$, $\lambda_2$ converges to zero while $\lambda_1$ diverges to infinity. In order for the quantity $\lambda_1 e_1^H p_{S,a}$ to converge, $e_1^H p_{S,a}$ should be zero, and thus

$$\tau_a = ix_a. \tag{91}$$

Note that the set of points satisfying (91) is line $\tau = ix$, which is equivalent to the $\tau'$-axis. Only the points satisfying (91) can appear in $S'$. Consider a point $p_{S,b}$ near line $\tau = ix$. The $p_{S',b}$ is expressed in $S'$ as

$$p_{S',b} = T(\theta_1) p_{S,b}. \tag{92}$$

Multiplying (92) by $E^H$ and using (85) and (86), it follows that

$$E^H p_{S',b} = \Lambda_{\theta_1} E^H p_{S,b}. \tag{93}$$

The vector $p_{S,b}$ can be represented as

$$p_{S,b} = \|p_{S,b}\| [\cos\theta_b, \sin\theta_b]^T. \tag{94}$$

As $\cos\theta_b$ tends to infinity, the point $p_{S,b}$ approaches line $\tau = ix$, and then $\|p_{S,b}\|$ reduces to zero. Thus the quantity $\lambda_1 e_1^H p_{S,b}$ becomes

$$\lim_{\cos\theta_b \to \infty} \lambda_1 e_1^H p_{S,b} = \frac{1}{\sqrt{2}} \lim_{v \to c} \|p_{S,b}\| \frac{(1+v/c)(1-v/c)}{1-(v/c)^2} = 0. \qquad (95)$$

From (83), (89), (90) and (95), we have

$$\begin{bmatrix} e_1^H p_{S',a} \\ e_2^H p_{S',a} \end{bmatrix} = \begin{bmatrix} \tau_a' - ix_a' \\ \tau_a' + ix_a' \end{bmatrix} = \begin{bmatrix} 0 \\ 0 \end{bmatrix}. \qquad (96)$$

Equation (96) leads to

$$\tau_a' = ix_a' \qquad (97)$$

$$\tau_a' = -ix_a'. \qquad (98)$$

Only the point $(\tau_a', x_a') = (0, 0)$ satisfies both (97) and (98). The point $(\tau', x') = (0, 0)$ is the origin of $S'$. Recall that according to (91), only the points on line $\tau = ix$ satisfy (95). Therefore, only the points having the relationship $\tau = ix$ in $S$ are all represented at the origin of $S'$. According to the relativity principle, only the points having the relationship $\tau' = -ix'$ in $S'$ are all represented at the origin of $S$.

## Figure Captions

Fig. 1. Coordinate systems with imaginary time axes.

Fig. 2. Comparison of the new coordinate system in CES and the Minkowski coordinate system in the real number space.

Fig. 3. Lines in $S'$ corresponding to lines in $S$ for $\alpha = 0$, $v/c$, $1$, and $\infty$.

Fig. 4. Coordinate system $S'^{(d')} = (\tau'^{(d')}, x'^{(d')})$.

Fig. 5. The set of points in $S'$ corresponding to line $\tau = \tau_a$ in $S$.

Fig. 6. Calculation of $d(p_{S',0})$ and $x_{cr}$.

Fig. 7. Propagation of $e_S(\tau_a, x_a)$ in $S$.

Fig. 8. Time intervals in $S'$ and $S$.

Fig. 9. The coordinates of $p_{S',a}$ in the real number space. (a) The $x'$-value of $p_{S',a}$. (b) The $ct'$-value of $p_{S',a}$.

Fig. 10. The set of points in $S'$ corresponding to line $ct = ct_a$.

Fig. 11. The set of points in $S'$ corresponding to line $x = x_a$.

Fig. 12. The set of points in $S$ corresponding to $\tau' = \tau_a'$ in $S'$ when $S' = \mathcal{S}'$.

Fig. 13. Movement of $O'^{(d')}$.

Fig. 14. Change of time into space.

Fig. 15. Calculation of the Hubble constant.

Fig. 16. Examples of twin paradox.

Fig. 17. Resolution of twin paradox.

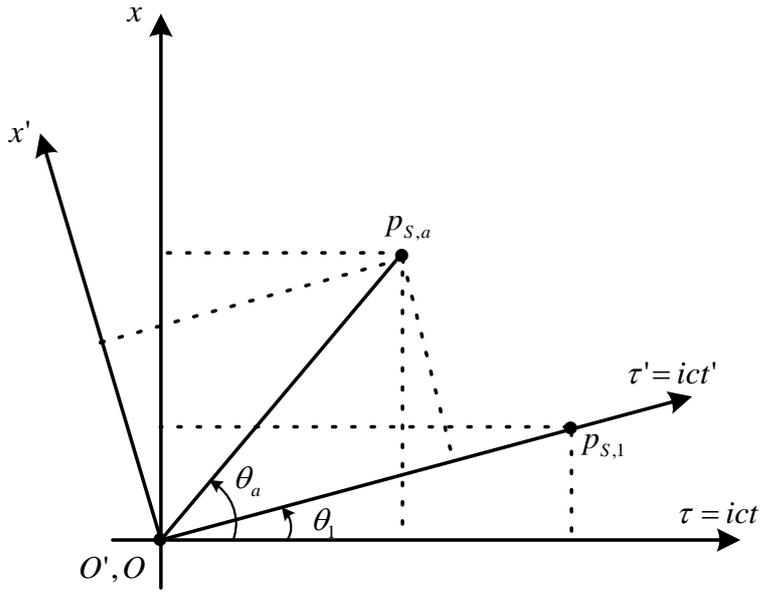

Fig. 1. Coordinate systems with imaginary time axes.

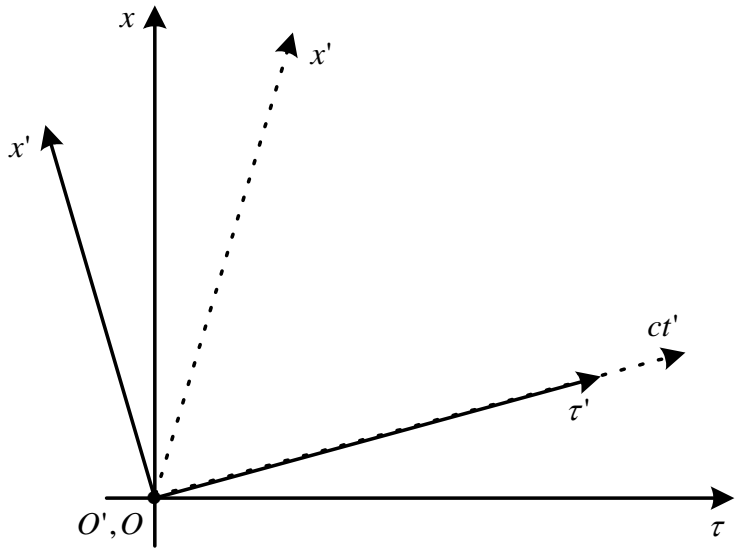

Fig. 2. Comparison of the new coordinate system in CES and the Minkowski coordinate system in the real number space.

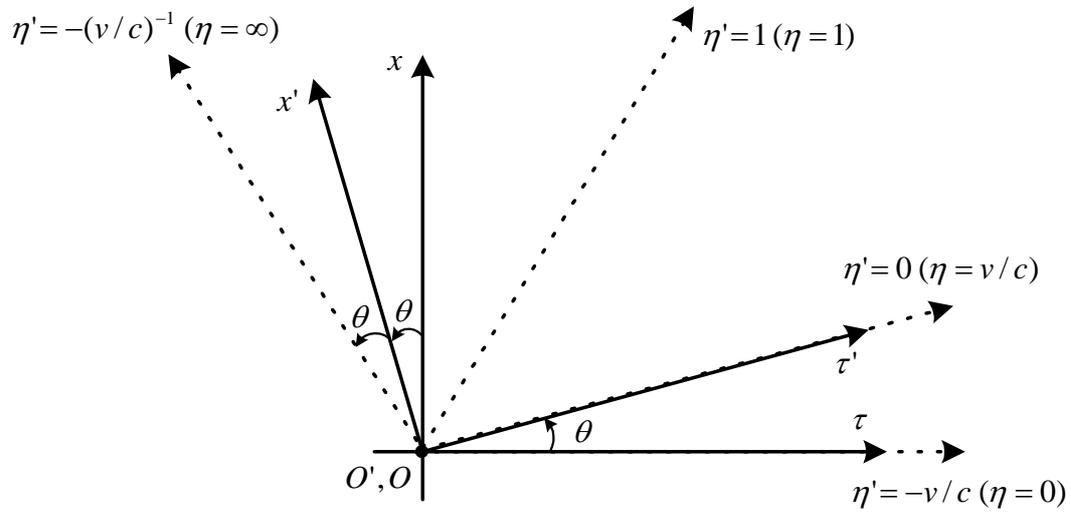

Fig. 3. Lines in $S'$ corresponding to lines in $S$ for $\alpha = 0$, $v/c$, $1$, and $\infty$.

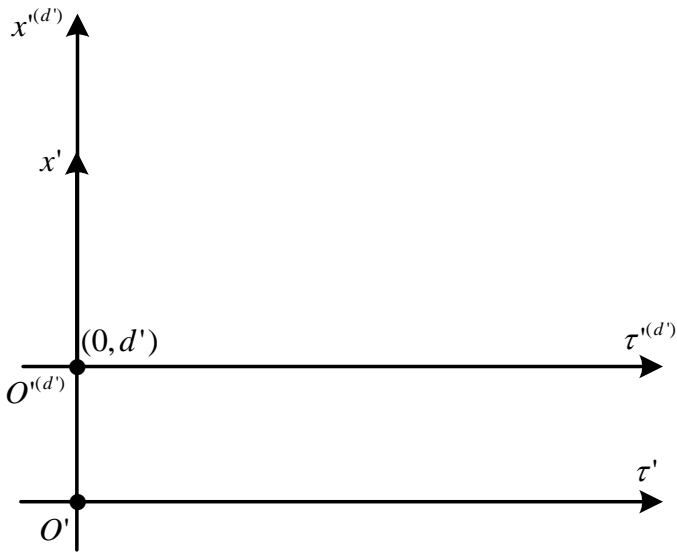

Fig. 4. Coordinate system $S'^{(d')} = (\tau'^{(d')}, x'^{(d')})$.

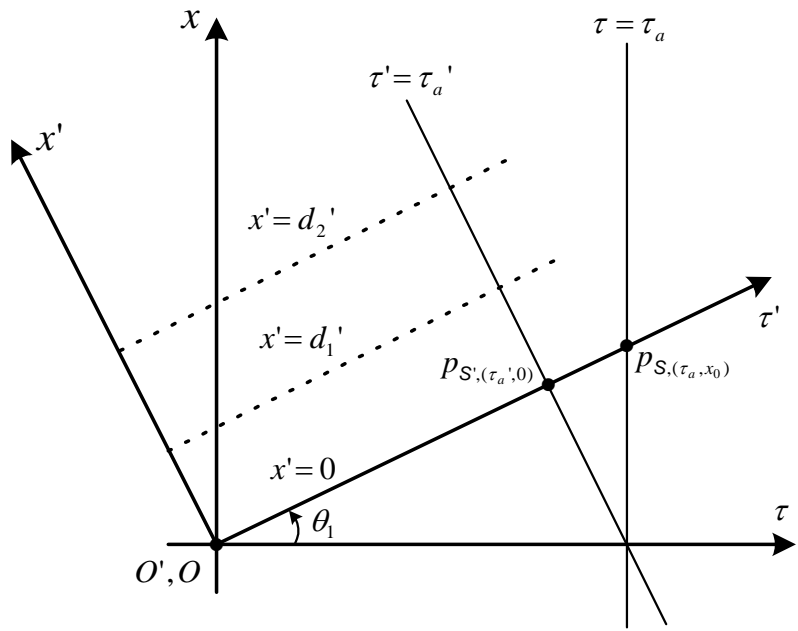

Fig. 5. The set of points in $S'$ corresponding to line $\tau = \tau_a$ in $S$.

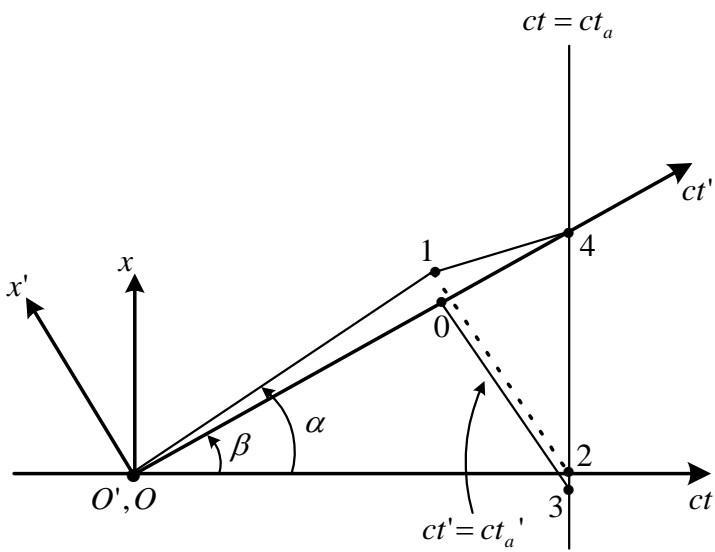

Fig. 6. Calculation of $d(p_{S',0})$ and $x_{cr}$.

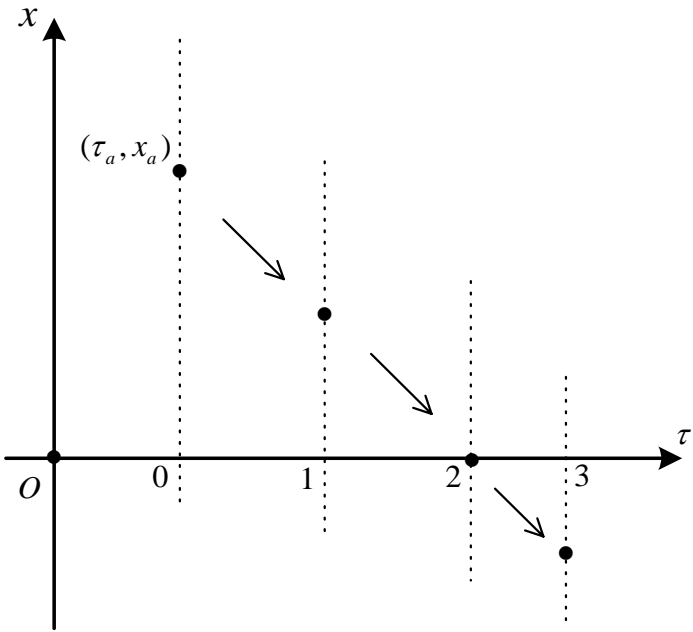

Fig. 7. Propagation of $e_S(\tau_a, x_a)$ in $S$.

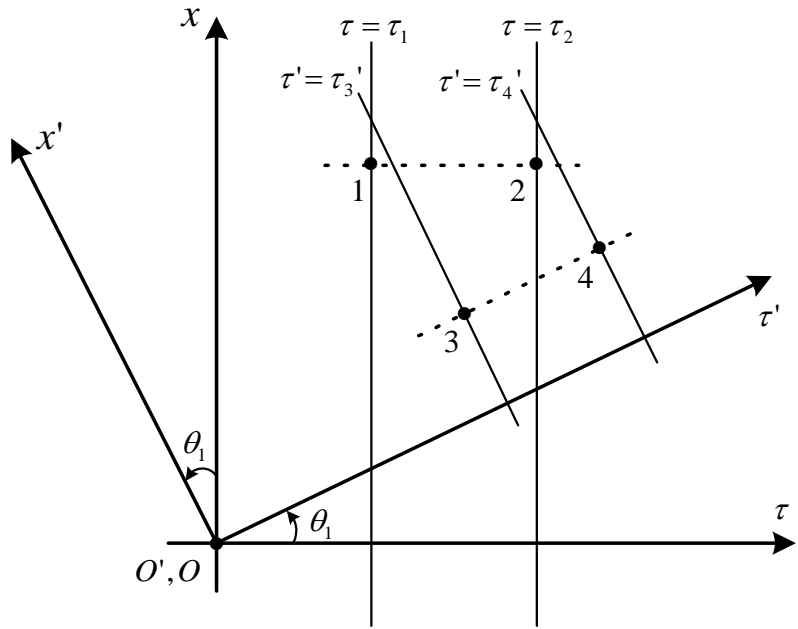

Fig. 8. Time intervals in $S'$ and $S$.

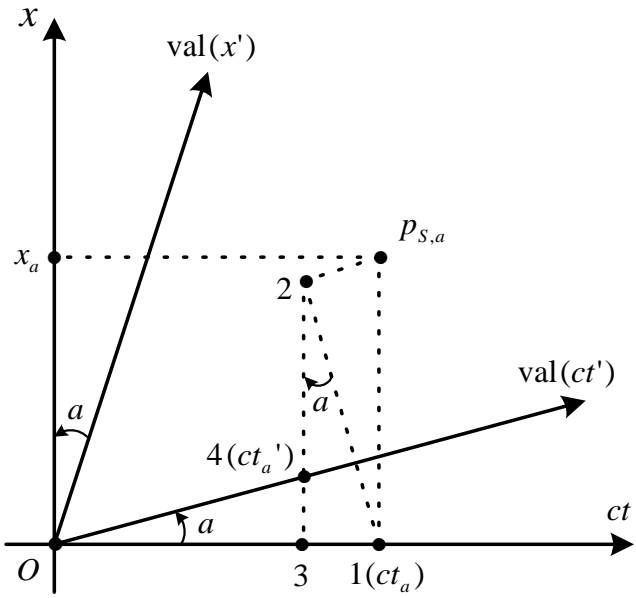

Fig. 9. The coordinates of $p_{S',a}$ in the real number space. (a) The $x'$-value of $p_{S',a}$.

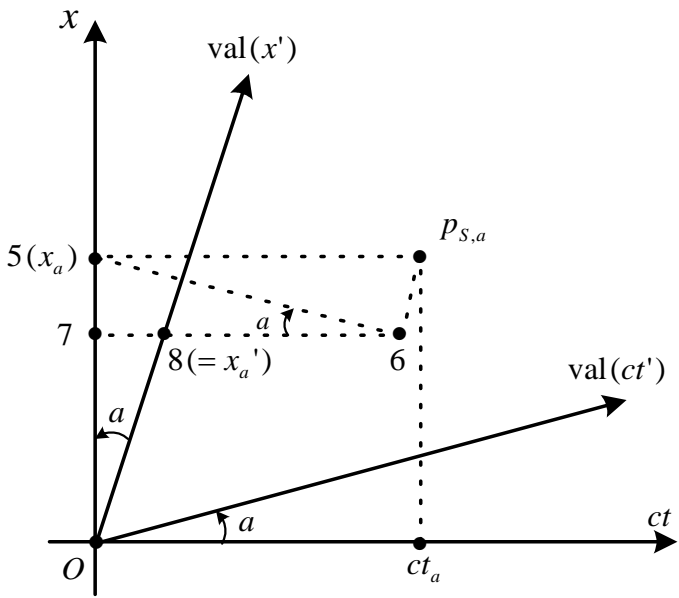

Fig. 9. The coordinates of $p_{S',a}$ in the real number space. (b) The $ct'$-value of $p_{S',a}$.

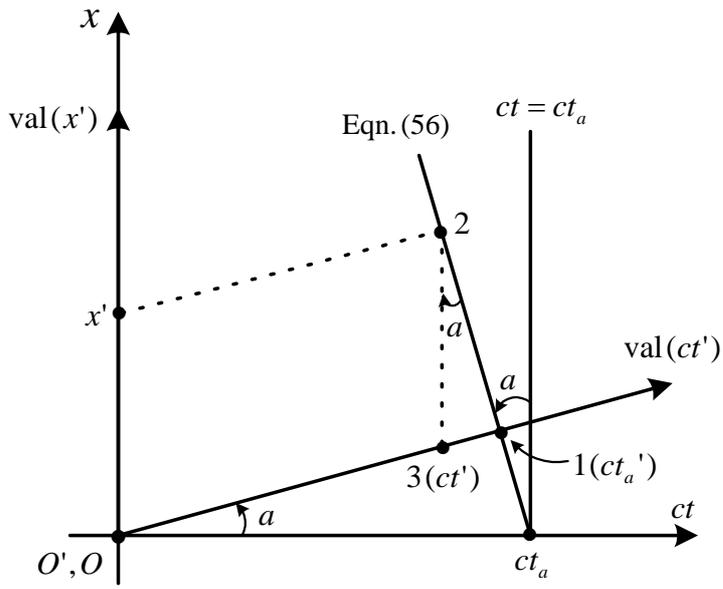

Fig. 10. The set of points in $S'$ corresponding to line $ct = ct_a$.

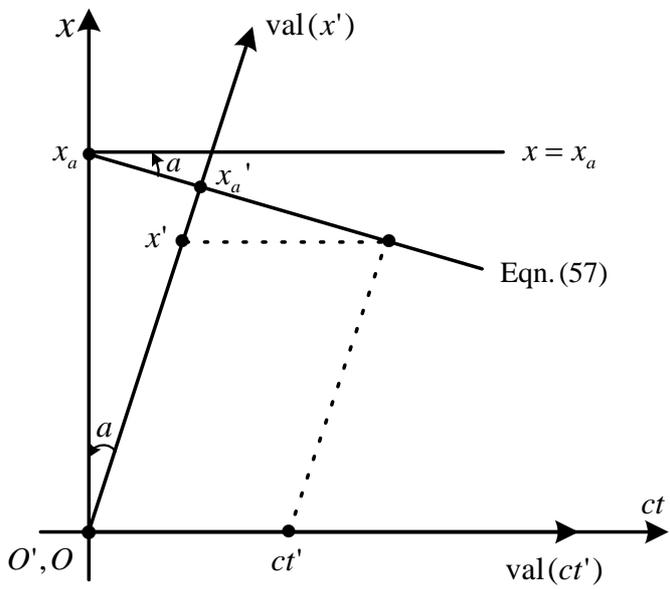

Fig. 11. The set of points in $S'$ corresponding to line $x = x_a$.

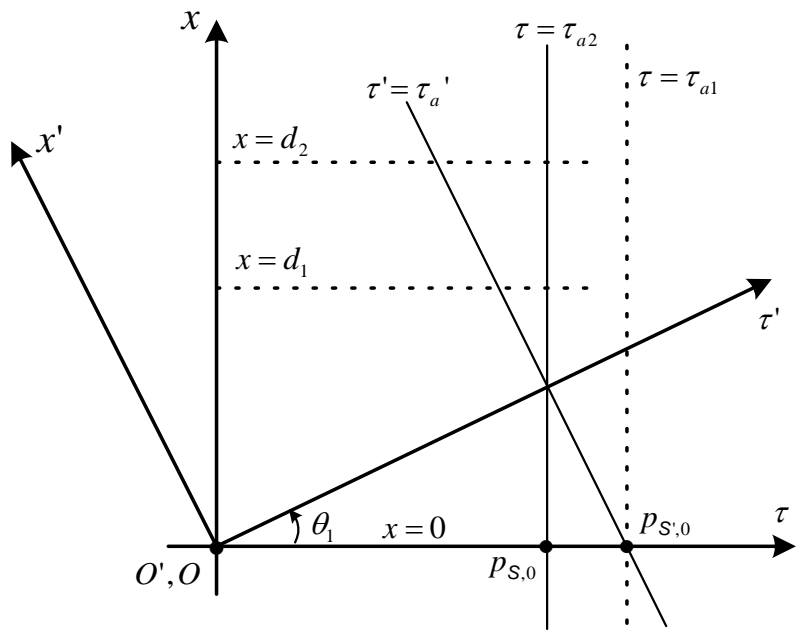

Fig. 12. The set of points in $S$ corresponding to $\tau' = \tau_a'$ in $S'$ when $S' = S'$.

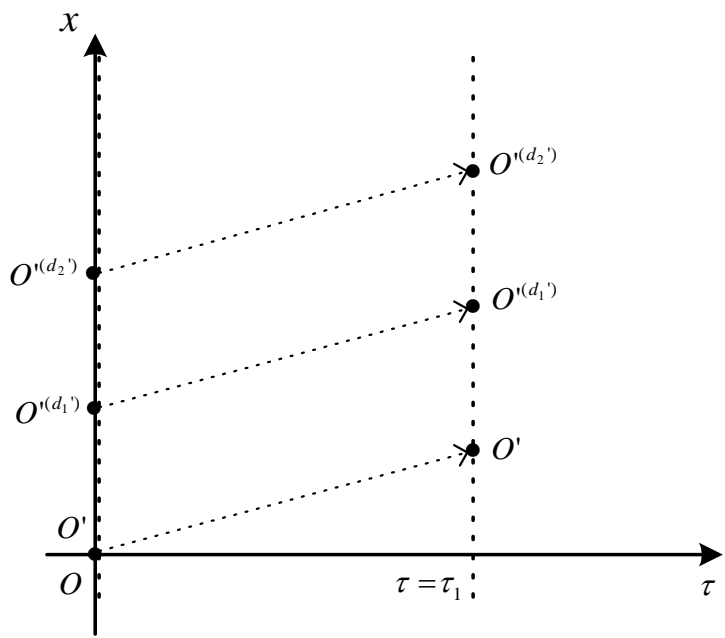

Fig. 13. Movement of $O'^{(d')}$.

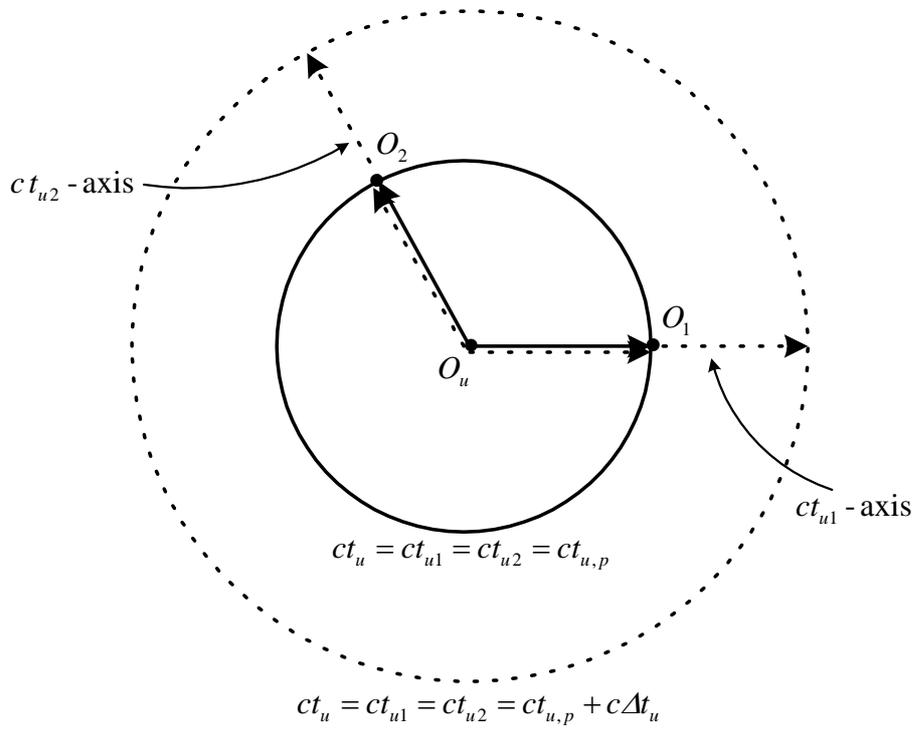

Fig. 14. Change of time into space.

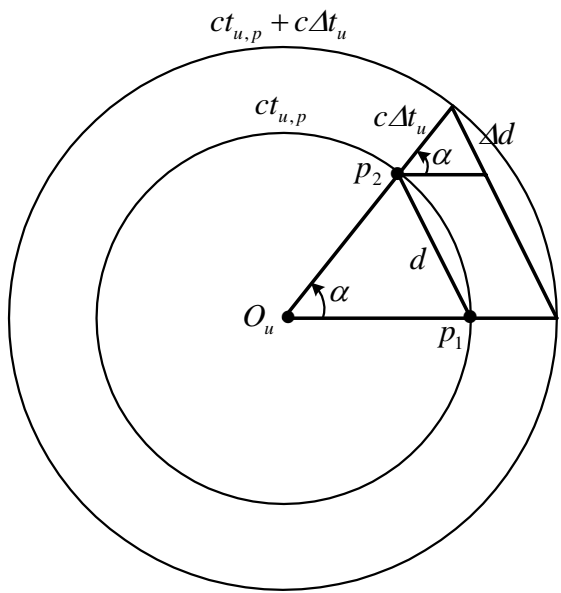

Fig. 15. Calculation of the Hubble constant.

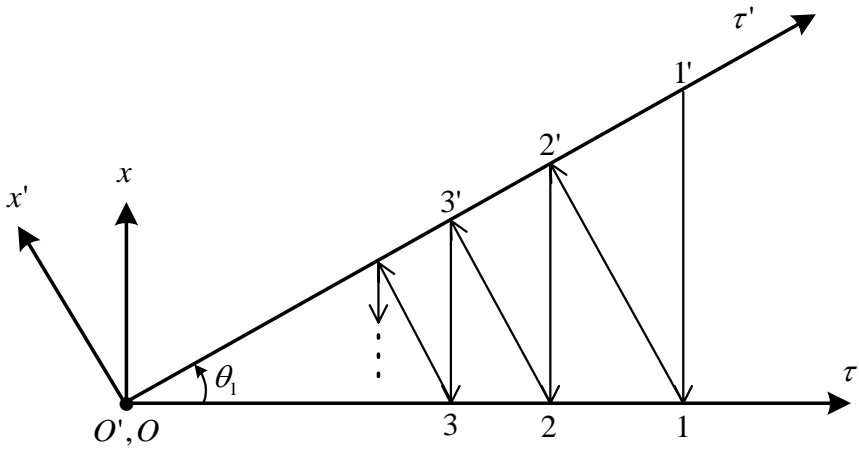

Fig. 16. Examples of twin paradox.

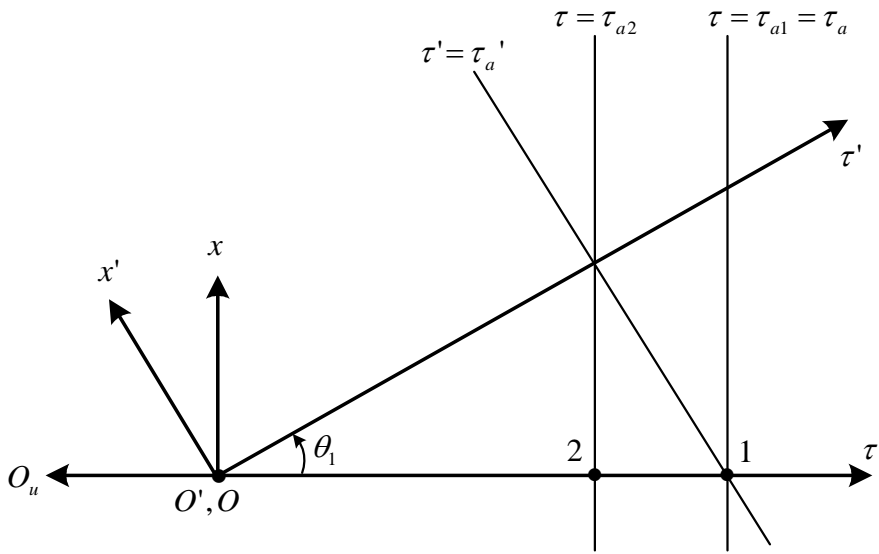

Fig. 17. Resolution of twin paradox.

35